\documentclass[submitting]{nst}

\usepackage{subfigure,dcolumn, supertabular, multirow, bigstrut}
\usepackage[T2A,T1]{fontenc}
\usepackage[russian,english]{babel}

\usepackage{listings, bm}
\begin{document}

\title{Probing the nucleon effective mass splitting with the light particle emission}\thanks{Supported by the National Natural Science Foundation of China (Nos. 12275359, 11875323, 11961141003, U2032145 and 11890712), the National Key R\&D Program of China under Grant (No. 2018YFA0404404), the Continuous Basic Scientific Research Project (No. WDJC-2019-13, BJ20002501), and the funding of China Institute of Atomic Energy (No. YZ222407001301). }

\author{Fang-Yuan Wang}
\affiliation{China Institute of Atomic Energy, Beijing 102413, China}
\author{Jun-Ping Yang}
\affiliation{China Institute of Atomic Energy, Beijing 102413, China}
\author{Xiang Chen}
\affiliation{China Institute of Atomic Energy, Beijing 102413, China}
\author{Ying Cui}
\affiliation{China Institute of Atomic Energy, Beijing 102413, China}
\author{Yong-Jia Wang}
\affiliation{School of Science, Huzhou University, Huzhou 313000, China}
\author{Zhi-Gang Xiao}
\affiliation{Department of Physics, Tsinghua University, Beijing, 100084, China}
\author{Zhu-Xia Li}
\affiliation{China Institute of Atomic Energy, Beijing 102413, China}
\author{Ying-Xun Zhang}
\email[Corresponding author, ]{Yingxun Zhang, zhyx@ciae.ac.cn} 
\affiliation{China Institute of Atomic Energy, Beijing 102413, China}
\affiliation{Guangxi Key Lab. Nucl. Phys. Techn., Guangxi Normal University, Guilin 541004, China}

\begin{abstract}
In this work, we first analyze the correlations among different nuclear matter parameters which are obtained by the 119 effective Skyrme interaction sets. The values of the correlation coefficients illustrate that the magnitude of the effective mass splitting is crucial for the tight constraints of the symmetry energy via HICs. Thus, the main object of this work is to investigate the impacts of the effective mass splitting on the heavy ion collisions observables. The $^{86}$Kr+$^{208}$Pb system at the beam energy ranging from 25A MeV to 200A MeV are simulated within the framework of the improved quantum molecular dynamics model (ImQMD-Sky) for this goal. Our calculations show that the slopes of the spectra of $\ln$[Y(n)/Y(p)] and $\ln$[Y(t)/Y($^3$He)], which are the logarithms of the neutron to proton yield ratios and triton to Helium-3 yield ratios, are directly related to the effective mass splitting and can be used to probe the effective mass splitting.
\end{abstract}

\keywords{effective mass splitting, symmetry energy, heavy ion collisions, Skyrme interaction}

\maketitle

\section{Introduction}
The nucleon effective mass $m_N^*$ is used to describe the motion of nucleons in a momentum-dependent potential equivalently with the motion of a quasi-nucleon of mass $m_N^*$ in a momentum-independent potential~\cite{BALi2018PPNP}. The isospin splitting of the nucleon effective mass means that the neutron effective mass is not equal to the proton effective mass, i.e., $m_n^*\ne m_p^*$, in isospin asymmetric nuclear matter. Both the effective mass and the effective mass splitting are important quantities in the isospin asymmetric nuclear equation of state, and serve as important microscopic inputs to the study of the thermal properties of the proto-neutron stars, such as thermal conductivity, specific heat, and neutrino reaction rates \cite{JEUKENNE76, SJOBERG76, Negele81, Meissner06, Steigman05, XLShang20, AngLi16}. Furthermore, the effective mass splitting is also important for improving the symmetry energy constraints \cite{Zhang2014, Coupland, ZhaoqingFeng,JunSu17,JunSu2020,ZhangF2020}.

There are a lot of efforts to constrain the effective mass splitting by using heavy ion collisions (HICs) \cite{Zhang2014, Coupland, ZhaoqingFeng,WJXie2013,JunSu17,WeiGF2020}, nucleon-nucleus optical potential \cite{BALi04, XiaoHuaLi, ChangXu}, the giant monopole resonance \cite{ZhenZhang2016,Kong2017,JunSu2020,JunXu20}. 
An interesting finding is that the effective mass splitting obtained by using the nucleon-nucleus optical potential analysis favours the $m_n^*>m_p^*$ \cite{BALi04}, while the HICs data support $m_n^*<m_p^*$ \cite{Zhang2014, Coupland, ZhaoqingFeng,JunSu17}. The possible reason for this discrepancy could be the matter that the different probes reflect the values of the effective mass splitting at different densities and momentum regions. To further understand the constraints of effective mass splitting by HICs, it is necessary to propose new probes for neutron-rich HICs and then compare the observables with experiments in the future.

Currently, the new generation rare isotope facilities or planned facilities, such as Heavy Ion Research Facility in Lanzhou (HIRFL/Lanzhou) \cite{zhouxh2018}, the Facility for Rare Isotope Beams at Michigan States University (FRIB/MSU) \cite{Ostroumov}, Radioactive Isotope Beam Factory (RIBF/RIKEN) \cite{Sakurai}, Rare isotope Accelerator complext for ON-line experiment (RAON/Korea) \cite{BHong}, and Beijing Isotope Separation On-line (BISOL/Beijing) \cite{liuwp2014}, can produce rare isotope beams from tens MeV per nucleon to hundreds MeV per nucleon for studying the dynamical evolution of neutron-rich nuclear system. Some important neutron-rich HICs experiments \cite{YanZhang14, Famiano07, Tsang09, Estee,LiLi2022} have been performed for extracting the information of the density dependence of the symmetry energy and effective mass splitting by comparing the data with the transport model simulations \cite{LWChen05, Tsang09}. 


Very recently, the experiment of $^{86}$Kr+$^{208}$Pb was performed with the Compact Spectrometer for Heavy IoN Experiment (CSHINE) \cite{Guan2021, Wang2021, wang2023, Diao2022}, which was installed at the final focal plane of the Radioactive Ion Beam Line in Lanzhou (RIBLL-I/HIRFL) \cite{Sun2003,WangHL2022}. The yield and kinetic variables of the A = 3 isobars, i.e., t and $^3$He, were measured. It provides an opportunity to constrain the symmetry energy \cite{zhang2005} at subsaturation density and to further check the capability of the transport models. In addition, in order to constrain the effective mass splitting by using the facilities in Lanzhou, the calculations are required to extend to the beam energy around 100-200A MeV, as this energy region has been found to be the optimal beam energy in previous calculations \cite{Rizzo2004,Zhang2014}.

In this paper, we will first analyze the correlations between different nuclear matter parameters to illustrate the significance of the investigation of effective mass splitting, and then investigate the impacts of effective mass splitting on the HICs  observables, such as neutron to proton yield ratios, triton to Helium-3 yield ratios, of $ ^{86} $Kr+$ ^{208} $Pb at the beam energy ranging from 25A MeV to 200A MeV by using the ImQMD-Sky model. 

\section{Theoretical Model}\label{sec:artwork}
In the improved quantum molecular dynamics model (ImQMD-Sky), each nucleon is represented by a Gaussian wave packet,
\begin{equation}
\label{phi1}
\phi_{k_{i0}}(\bm{r}_i)=\frac{1}{(2\pi\sigma_r^2)^{3/4}} \exp\left[-\frac{(\bm{r}_i-\bm{r}_{i0})^2}{4\sigma_r^2}+{\rm i}(\bm{r}_i-\bm{r}_{i0})\cdot \bm{k}_{i0}\right]
\end{equation}
here, $\sigma_r$ and $\bm{r}_{i0}$ are the width and centroid of wave packet, respectively. The $k_{i0}$ in the subscript corresponds to the state of the $i$th nucleon. For an $N$-body system, the system wave function is assumed to be a direct product of $N$ coherent states,
\begin{eqnarray}
\Psi(\bm{r}_1, \cdots  ,\bm{r}_N)=\phi_{k_1}(\bm{r}_1)\phi_{k_2}(\bm{r}_2) \cdots \phi_{k_N}(\bm{r}_N).
\end{eqnarray}
By using the Wigner transformation, the phase space density distribution of the system can be obtained as
%
\begin{equation} \label{f1qmd}
\begin{aligned}
    f_N&(\bm{r}_1, \cdots  ,\bm{r}_N; \bm{p}_1, \cdots  ,\bm{p}_N) \\
    &= \Pi_{i} \frac{1}{(\pi\hbar)^{3}}\exp\left[-\frac{(\bm{r}_i-\bm{r}_{i0})^2}{2\sigma_r^2}-\frac{(\bm{p}_i-{\bm{p}_{i0}})^2}{2\sigma_p^2}\right],\vspace*{0.5mm}
\end{aligned}
\end{equation}
where $\sigma_r\sigma_p=\hbar/2$. $\bm{r}_i$ and $\bm{p}_i$ are the position and momentum of the $i$th nucleon. The Hamiltonian of the nucleonic part is calculated as
%
\begin{equation} \label{Hamiltonian}
\begin{aligned}
    H &= \langle \Psi|\hat{T}+\hat{U}|\Psi\rangle \\
    & \equiv \sum_{i}\int \frac{\hat{\bm{p}}_i^2}{2m}f_i(\bm{r},\bm{p})\text{d}^3r\text{d}^3p \\
    & \quad + \sum_{i< j}\int \hat{v}^{ij}_{sky} f_i(\bm{r},\bm{p})f_j(\bm{r}',\bm{p}')\text{d}^3 r\text{d}^3p\text{d}^3 r'\text{d}^3p'\\
    & = \sum_{i}\left(\frac{p_{i0}^2}{2m}+C(\sigma_r)\right)+\int u_{sky} \text{d}^3 r.
\end{aligned}
\end{equation}
$C(\sigma)=\frac{1}{2m}\frac{3\hbar^2}{4\sigma_r^2}$ is the contribution of the width of the wave packet in the kinetic energy term. 
$u_{sky}$ is the potential energy density in coordinate space.

For the nucleonic potential part, the Skyrme type nucleonic potential energy density without the spin-orbit term is used,
\begin{equation}
    u_\text{sky}=u_\text{loc}+u_\text{md}.
\end{equation}
The local potential energy density is,
%
\begin{eqnarray} \label{eq:edfimqmd}
\begin{aligned}
    u_\text{loc} &= \frac{\alpha}{2}\frac{\rho^2}{\rho_0} +\frac{\beta}{\eta+1}\frac{\rho^{\eta+1}}{\rho_0^\eta}+\frac{g_{sur}}{2\rho_0 }(\nabla \rho)^2 \\
    & \quad +\frac{g_{sur,iso}}{\rho_0}[\nabla(\rho_n-\rho_p)]^2 \\
    & \quad +A_\text{sym}\frac{\rho^2}{\rho_0}\delta^2+B_\text{sym}\frac{\rho^{\eta+1}}{\rho_0^\eta}\delta^2.
\end{aligned}
\end{eqnarray}
$\rho=\rho_n+\rho_p$ is the nucleon density and $\delta=(\rho_n-\rho_p)/\rho$ is the isospin asymmetry. The $\alpha$ is the parameter related to the two-body term, $\beta$ and $\eta$ are related to the three-body term, $g_{sur}$ and $g_{sur,iso}$ are related to the surface terms, $A_\text{sym}$ and $B_\text{sym}$ are the coefficients in the symmetry potential and come from the two- and the three-body interaction terms \cite{Zhang20FOP}. Their values can be obtained from the standard Skyrme interaction.

The non-local potential energy density or the momentum dependent interaction term, i.e., $u_{md}$, is also taken as a Skyrme-type momentum dependent energy density functional.  It is obtained based on its interaction form $\delta (\bm r_1-\bm r_2 ) (\bm p_1-\bm p_2 )^2$ \cite{Skyrme1956}, i.e.,
%
\begin{eqnarray} \label{eq:mdimqmd}
\begin{aligned}
    u_{md} & = C_0\sum_{ij}\int \text{d}^3p\text{d}^3p' f_i(\bm r,\bm p)f_j(\bm r,\bm p')(\bm p-\bm p')^2 \\
    & \quad + D_0\sum_{ij\in n}\int \text{d}^3 p \text{d}^3 p'f_i(\bm r,\bm p) f_j(\bm r,\bm p')(\bm p-\bm p')^2 \\
    & \quad + D_0\sum_{ij\in p}\int \text{d}^3p \text{d}^3p' f_i(\bm r,\bm p)f_j(\bm r,\bm p')(\bm p-\bm p')^2.
\end{aligned}
\end{eqnarray}
$C_0$ and $D_0$ are the parameters related to the momentum dependent interaction, and they are related to the standard Skyrme interaction as follows,
\begin{equation}
\begin{aligned}
    C_0 &= \frac{1}{16\hbar^2}\left[t_1(2+x_1)+t_2(2+x_2) \right] \\
    D_0 &= \frac{1}{16\hbar^2}\left[t_2(2x_2+1)-t_1(2x_1+1) \right].
\end{aligned}
\end{equation}
More details about it can be found in Ref. \cite{JPYang21}. The parameters in Eq. \eqref{eq:edfimqmd} and \eqref{eq:mdimqmd} are obtained from the standard Skyrme interaction parameters as in Refs. \cite{YXZhang06,YXZhang20}. The Coulomb term is treated by the standard method in QMD type models.




The initialization is prepared as the same as in Ref. \cite{Zhang20FOP}. 
The centroids of the wave packet for neutrons and protons are sampled within an empirical radius of the neutron and proton \cite{Zhang20FOP}. After the positions of all nucleons are finally prepared, the density distribution will be known. The momenta of the nucleons are sampled according to the local density approach. 

Here, it should be noted that the effects of the width of the wave packet in the sampling of the momentum in the initialization are considered in this work. Usually, $C(\sigma)$ was omitted in the QMD type models for the study of intermediate-high energy HICs since it has no effect on the equation of motion and its correction to the initial momentum is relatively small. However, it cannot be neglected, especially for studying low energy reactions. The reason is that $C(\sigma)$ in the kinetic energy term will reach about 25\% of the fermi energy at normal density, $\sim$35 MeV. For example, $C(\sigma)$ will be 8.97 MeV when the width of the wave packet takes a typical value, i.e., $\sigma_r$=1.32 fm. For the expected values of the momentum of the nucleons which are sampled in the calculations, the width of the wave packet has no direct effects since $\langle \phi_i|\bm{p}|\phi_i\rangle=\bm{p}_{i0}$. To satisfy the requirements of reasonably describing the binding energy of initial nuclei with Gaussian wave packet\cite{ZhangF2020,Hermann2022,Xujun2016}, the sampled $\bm{p}_{i0}$ should be shrunk to a smaller value than that without considering the width of wave packet. %

\section{Results and Discussion} \label{results}
To see the importance of the effective mass splitting on the symmetry energy constraints, we will first analyze the correlations between different nuclear matter parameters. Then, the influence of effective mass splitting on the HICs is presented and discussed.
\subsection{The nuclear matter parameters and its correlations}
For the Skyrme effective interaction used in this work, the corresponding isospin asymmetric equation of state for cold nuclear matter reads,
%
\begin{equation} \label{eossky}
\begin{aligned}
    E/A &= \frac{3\hbar^2}{10m}\left(\frac{3\pi^2}{2}\rho\right)^{2/3} \\
    & \quad +\frac{\alpha}{2}\frac{\rho}{\rho_{0}}+\frac{\beta}{\eta+1}\frac{\rho^{\eta}} 
        {\rho^{\eta}_{0}}+g_{\rho\tau}\frac{\rho^{5/3}}{\rho_{0}^{5/3}} + S(\rho)\delta^2,
\end{aligned}
\end{equation}
where the density dependence of the symmetry energy $S(\rho)$ is,
%
\begin{equation} \label{SE-skyrme}
\begin{aligned}
    S(\rho) & = \frac{\hbar^2}{6m}\left(\frac{3\pi^2\rho}{2}\right)^{2/3}+A_\text{sym}\frac{\rho}{\rho_{0}} \\
    & \quad + B_\text{sym}\left(\frac{\rho}{\rho_{0}}\right)^{\eta}+C_\text{sym}(m_s^*,m_v^*)\left(\frac{\rho}{\rho_{0}}\right)^{5/3}.
\end{aligned}
\end{equation}
The terms of $g_{\rho\tau}$ in Eq.~(\ref{eossky}) and $C_\text{sym}$ in Eq. (\ref{SE-skyrme}) come from the energy density functional of the Skyrme-type momentum dependent interaction and their relations to the standard Skyrme interaction can be found in Refs. \cite{Zhang2006}. The pressure in the nuclear fluid can be calculated as follows:
\begin{equation}
    P=\rho^2\frac{\partial E/A(\rho,\delta)}{\partial \rho}.
\end{equation}
The saturation density $\rho_0$ for symmetric nuclear matter is obtained by
\begin{equation}
\label{rho0}
    P=\rho_0^2\left( \frac{d}{d\rho}\frac{E}{A}(\rho,\delta=0)\right)|_{\rho=\rho_0}=0.
\end{equation}

Correspondingly, the nuclear matter parameters at saturation density can be obtained. For example, the binding energy $E_0$ and the incompressibility $K_0$ are,
\begin{eqnarray}
E_0&=&E/A(\rho_0),\\
K_0&=&9\rho_0^2\frac{\partial^2 E/A}{\partial \rho^2}|_{\rho_0}.
\end{eqnarray}
The symmetry energy coefficient $S_0$ and the slope of the symmetry energy $L$ are
\begin{eqnarray}
    S_0&=&S(\rho_0),\\
    L&=&3\rho_0\frac{\partial S(\rho)}{\partial \rho}|_{\rho_0}.
\end{eqnarray}

The neutron/proton effective mass is obtained from the neutron to proton potential according to,
\begin{equation}
\label{effm}
    \frac{m}{m_q^*} = 1+\frac{m}{p}\frac{\partial V_q}{\partial p} , \quad q=n,p.   
\end{equation}
$V_q$ is the single particle potential for neutron or proton and the form of $V_q$ can be found in appendix~\ref{Vq}. For Skyrme interaction, the neutron/proton effective mass will is,
\begin{equation}
\label{effmq}
    \frac{m}{m_q^*}=1+4mC_0\rho+4mD_0\rho_q. 
\end{equation}
The isoscalar effective mass $m_s^*$ can be obtained at $\rho_q=\rho/2$ from Eq. \eqref{effmq}, and the isovector effective mass $m_v^*$ can be obtained at $\rho_q=0$ which represents the neutron(proton) effective mass in pure proton(neutron) matter as in Refs. \cite{chabnate,ZhenZhang2016}. They are,
\begin{eqnarray}
    \frac{m}{m_s^*}&=& 1+4m\left(C_0+\frac{D_0}{2}\right)\rho ,\\
    \frac{m}{m_v^*}&=& 1+4mC_0\rho .
\end{eqnarray}
By using $m_s^*$ and $m_v^*$, the effective mass splitting $\delta m_{np}^*=(m_n^*-m_p^*)/m$ can be expressed as
\begin{equation}
\label{delta_mnp}
   \delta m_{np}^*=\frac{m_n^*-m_p^*}{m}=2\frac{m_s^*}{m}\sum_{n=1}^\infty \left(\frac{m_s^*-m_v^*}{m_v^*}\right)^{2n-1}\delta^{2n-1},
\end{equation}
as in Ref. \cite{ZhenZhang2016}. As described in Eq. \eqref{delta_mnp}, the exact value of $\delta m_{np}^*=(m_n^*-m_p^*)/m$ depends on the expansion and the isospin asymmetry of the system $\delta$. To avoid the dependence on the expansion and $\delta$, we define a quantity $f_I$
\begin{equation}
    f_I=\frac{1}{2\delta} \left( \frac{m}{m_n^*}-\frac{m}{m_p^*} \right) = \frac{m}{m_s^*}-\frac{m}{m_v^*},
\end{equation}
to describe the isospin effective mass splitting, which has the opposite sign with $\delta m_{np}^*$.

Since the parameters of the nuclear matter mentioned above are obtained from the same energy density functional, one can expect there is a correlation among them. For example, as shown in Eq. \eqref{SE-skyrme}, the $S(\rho)$ depends on the two-body, three-body, and momentum-dependent interaction terms. These three terms are correlated to the $E_0$, $K_0$ and $m_s^*$ \cite{chabnate}, and also to the $S_0$, $L$, and $m_v^*$~\cite{YXZhang20}. However, the correlation strength depends on the effective Skyrme interaction parameter set one used~\cite{chabnate}.

To describe the correlation among the different nuclear matter parameters with less bias, one can calculate the linear correlation coefficient $C_{AB} $ between the nuclear matter parameters $A$ and $B$ based on the current knowledge of the nuclear matter parameters ~\cite{YXZhang20}, i.e., 
\begin{equation}
\label{MP-Criteria}
\begin{aligned}
    200 \text{ MeV} \leqslant K_0 & \leqslant 280 \text{ MeV}, \\
     25 \text{ MeV} \leqslant S_0 & \leqslant  35 \text{ MeV}, \\
     30 \text{ MeV} \leqslant  L & \leqslant 120 \text{ MeV}, \\
    0.6             \leqslant m_s^*/&m \leqslant 1.0,       \\
   -0.5             \leqslant f_I & \leqslant 0.4.
\end{aligned}
\end{equation}
The quantity $A$ or $B$=$\{\rho_0, E_0, K_0, S_0, L, m_s^*, m_v^*\}$, the correlation coefficient $C_{AB}$ is calculated as follows, 
\begin{equation}
\begin{aligned}
    C_{AB} &= \frac{ \text{cov} ( A, B ) }{ \sigma ( A ) \sigma ( B ) }, \\
    \text{cov} ( A, B ) &= \frac{1}{N-1} \sum_i ( A_i - \langle A \rangle ) ( B_i - \langle B 
        \rangle ), \\
    \sigma ( X ) &= \sqrt{ \frac{1}{N-1} \sum_i ( X_i - \langle X \rangle )^2 }, \quad X = A, B \\
    \langle X \rangle &= \frac{1}{N} \sum_i X_i, \quad i = 1, N.
\end{aligned}
\end{equation}
$\text{cov} ( A, B )$ is the covariance between $A$ and $B$, $\sigma(X)$ is the standard deviation of $X$. $\langle X \rangle$ means the average values obtained from $N=119$ standard Skyrme parameter sets, which are selected according to the criteria in Eq.(\ref{MP-Criteria}).

\begin{figure}[!ht]
    \centering
    \includegraphics[width=\linewidth]{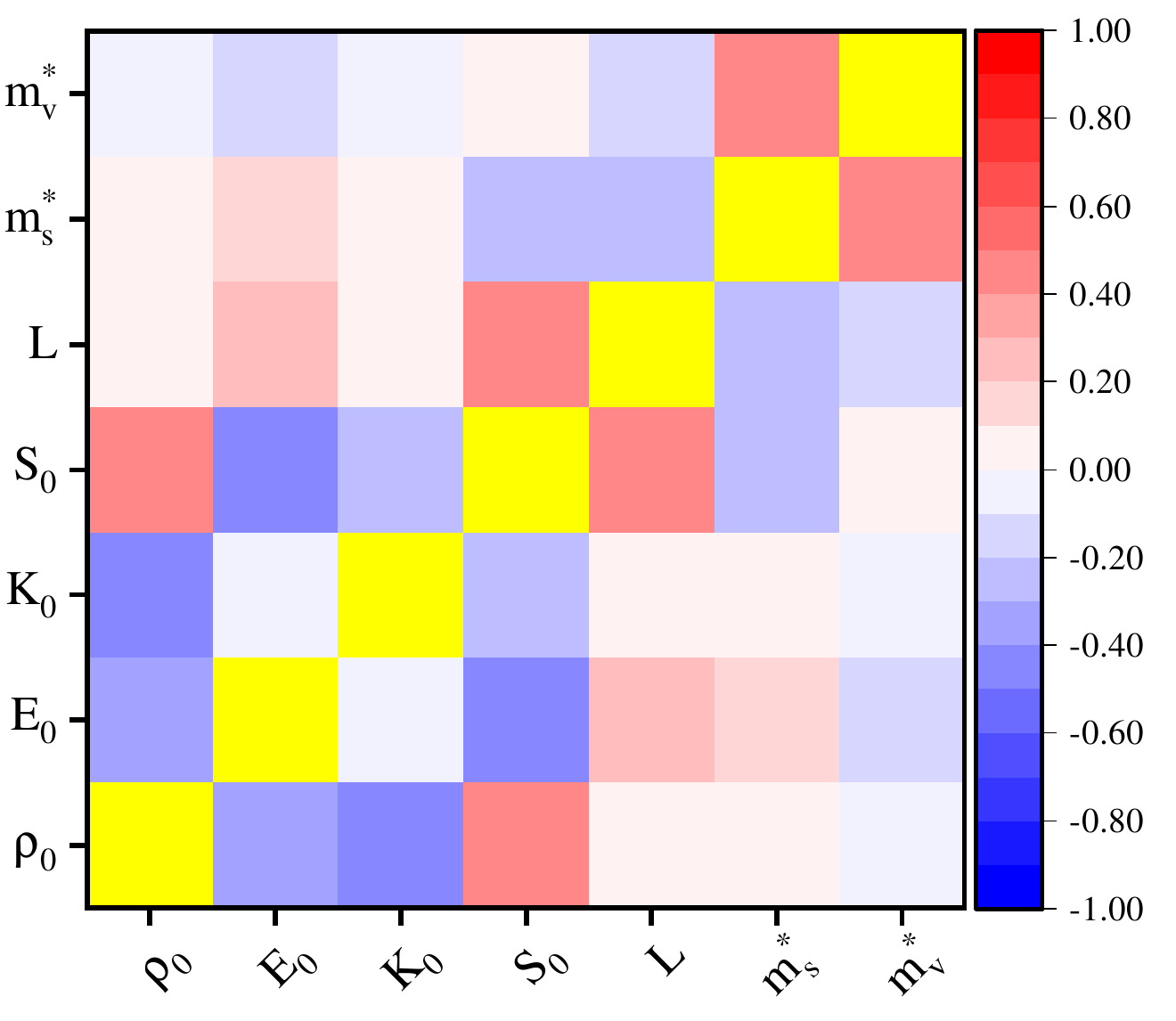}
    \caption{The correlation coefficients among the different nuclear matter parameter pairs. 
    }
    \label{fig2}
\end{figure}

The values of these parameters are listed in Table~\ref{tbl401}, and the correlation coefficient $C_{AB}$ is illustrated in  Figure~\ref{fig2}. 
The positive value of $C_{AB}$ reflects the positive linear correlation, while the negative value means the negative linear correlation. As one can see, the correlations exist among different nuclear matter parameters. In more detail, the correlation between $ S_0 \text{ and } \rho_0 $, $ L \text{ and } S_0 $, $ m_v^* \text{ and } m_s^* $, $ K_0 \text{ and } \rho_0 $, $ S_0 \text{ and } E_0 $, are stronger than other nuclear matter parameter pairs. The `strang' correlation between $\rho_0$ and $S_0$ can be understood as follows. As we know, $\rho_0$ is determined by Eq.(\ref{rho0}), which is related to the parameters $\alpha$, $\beta$, $\eta$ and $g_{\rho\tau}$, or related to nuclear matter parameters as presented in Eq.(5) of Ref.\cite{YXZhang20}. These correlations mean that tight constraints on the density dependence of the symmetry energy by using HICs need to know not only the information of $S_0$ and $L$, but also the $m_s^*$ and $m_v^*$ (or the effective mass splitting). 

\begin{center} \tablecaption{Nuclear matter parameters calculated from 119 
    Skyrme interaction sets.} \label{tbl401}
\tablefirsthead{
    \hline
    Para. & \multicolumn{1}{c}{$\rho_0$} & \multicolumn{1}{c}{$E_0$} & 
    \multicolumn{1}{c}{$K_0$} & \multicolumn{1}{c}{$S_0$} & 
    \multicolumn{1}{c}{$L$} & \multicolumn{1}{c}{$m_s^*$} & 
    \multicolumn{1}{c}{$m_v^*$} \\ \hline }
\tablehead{
    \multicolumn{8}{c}{Table \ref{tbl401}. (Continued.)} \\ \hline
    Para. & \multicolumn{1}{c}{$\rho_0$} & \multicolumn{1}{c}{$E_0$} & 
    \multicolumn{1}{c}{$K_0$} & \multicolumn{1}{c}{$S_0$} & 
    \multicolumn{1}{c}{$L$} & \multicolumn{1}{c}{$m_s^*$} & 
    \multicolumn{1}{c}{$m_v^*$} \\ \hline }
\tabletail{
    \hline
    \multicolumn{8}{r}{\small\sl continued on the next column} \\}
\tablelasttail{\hline}
\begin{supertabular}{c*{7}{r}}
BSk9 & 0.159  & -15.90  & 231.56  & 30.00  & 39.90  & 0.80  & 0.91  \\
BSk10 & 0.159  & -15.89  & 238.17  & 29.98  & 37.34  & 0.92  & 0.81  \\
BSk11 & 0.159  & -15.84  & 239.03  & 30.04  & 38.34  & 0.92  & 0.82  \\
BSk12 & 0.159  & -15.84  & 238.99  & 30.04  & 37.98  & 0.92  & 0.82  \\
BSk13 & 0.159  & -15.84  & 239.02  & 30.04  & 38.81  & 0.92  & 0.82  \\
BSk14 & 0.159  & -15.83  & 240.29  & 30.04  & 43.92  & 0.80  & 0.78  \\
BSk15 & 0.159  & -16.02  & 241.70  & 30.00  & 33.62  & 0.80  & 0.77  \\
BSk16 & 0.159  & -16.03  & 242.63  & 30.03  & 34.83  & 0.80  & 0.78  \\
BSk17 & 0.159  & -16.03  & 242.65  & 30.03  & 36.25  & 0.80  & 0.78  \\
FPLyon & 0.162  & -15.90  & 217.20  & 30.94  & 42.78  & 0.84  & 0.97  \\
Gs & 0.158  & -15.57  & 238.13  & 31.46  & 94.32  & 0.78  & 0.68  \\
KDE & 0.164  & -15.97  & 223.13  & 31.93  & 41.44  & 0.76  & 0.86  \\
KDE0v & 0.161  & -16.08  & 229.01  & 32.99  & 45.22  & 0.72  & 0.77  \\
KDE0v1 & 0.165  & -16.21  & 228.33  & 34.62  & 54.74  & 0.74  & 0.81  \\
LNS & 0.175  & -15.29  & 211.47  & 33.48  & 61.55  & 0.83  & 0.73  \\
MSk1 & 0.157  & -15.81  & 232.62  & 29.96  & 34.05  & 1.00  & 1.00  \\
MSL0 & 0.160  & -15.86  & 230.26  & 29.98  & 59.97  & 0.80  & 0.70  \\
NRAPR & 0.161  & -15.83  & 226.52  & 32.84  & 59.73  & 0.69  & 0.60  \\
RATP & 0.160  & -16.02  & 239.84  & 29.27  & 32.41  & 0.67  & 0.56  \\
Rs & 0.158  & -15.57  & 237.94  & 30.63  & 85.88  & 0.78  & 0.68  \\
Sefm074 & 0.160  & -15.79  & 239.17  & 33.33  & 88.59  & 0.74  & 0.63  \\
Sefm081 & 0.161  & -15.66  & 237.24  & 30.79  & 79.48  & 0.81  & 0.68  \\
Sefm09 & 0.161  & -15.53  & 240.24  & 27.80  & 70.05  & 0.90  & 0.75  \\
SGI & 0.154  & -15.87  & 260.52  & 28.27  & 63.76  & 0.61  & 0.58  \\
SGII & 0.158  & -15.57  & 213.95  & 26.81  & 37.70  & 0.79  & 0.67  \\
SKa & 0.155  & -15.97  & 262.15  & 32.86  & 74.56  & 0.61  & 0.52  \\
Ska25s20 & 0.161  & -16.05  & 221.45  & 33.83  & 63.90  & 0.98  & 0.98  \\
SkI2 & 0.158  & -15.75  & 241.98  & 33.47  & 104.71  & 0.68  & 0.80  \\
SkI4 & 0.160  & -15.92  & 247.64  & 29.48  & 60.36  & 0.65  & 0.80  \\
SkI6 & 0.159  & -15.90  & 248.40  & 30.07  & 59.67  & 0.64  & 0.80  \\
SkM & 0.160  & -15.75  & 216.00  & 30.72  & 49.39  & 0.79  & 0.66  \\
SkMs & 0.160  & -15.75  & 216.00  & 30.01  & 45.84  & 0.79  & 0.65  \\
SkMP & 0.157  & -15.54  & 230.74  & 29.88  & 70.33  & 0.65  & 0.59  \\
SkO & 0.160  & -15.81  & 222.41  & 31.90  & 79.00  & 0.90  & 0.85  \\
SkOp & 0.160  & -15.73  & 221.94  & 31.92  & 68.92  & 0.90  & 0.87  \\
SKRA & 0.159  & -15.75  & 216.08  & 31.28  & 53.07  & 0.75  & 0.63  \\
SkS1 & 0.161  & -15.84  & 227.93  & 28.74  & 30.65  & 0.86  & 0.64  \\
SkSC14 & 0.161  & -15.90  & 235.96  & 30.02  & 33.11  & 1.00  & 1.00  \\
SkT1 & 0.161  & -15.96  & 236.10  & 32.02  & 56.22  & 1.00  & 1.00  \\
SkT1s & 0.162  & -15.95  & 239.83  & 32.23  & 56.27  & 1.00  & 1.00  \\
SkT1a & 0.161  & -15.96  & 236.10  & 32.02  & 56.22  & 1.00  & 1.00  \\
SkT2 & 0.161  & -15.92  & 235.66  & 32.00  & 56.20  & 1.00  & 1.00  \\
SkT2a & 0.161  & -15.92  & 235.66  & 32.00  & 56.20  & 1.00  & 1.00  \\
SkT3 & 0.161  & -15.92  & 235.70  & 31.50  & 55.35  & 1.00  & 1.00  \\
SkT3a & 0.161  & -15.92  & 235.70  & 31.50  & 55.35  & 1.00  & 1.00  \\
SkT6 & 0.161  & -15.94  & 236.21  & 29.97  & 30.85  & 1.00  & 1.00  \\
SkT6a & 0.161  & -15.94  & 236.21  & 29.97  & 30.85  & 1.00  & 1.00  \\
SkT7 & 0.161  & -15.92  & 236.45  & 29.55  & 31.08  & 0.83  & 0.71  \\
SkT7a & 0.161  & -15.92  & 236.45  & 29.55  & 31.08  & 0.83  & 0.71  \\
SkT8 & 0.161  & -15.92  & 236.40  & 29.94  & 33.69  & 0.83  & 0.83  \\
SkT8a & 0.161  & -15.92  & 236.40  & 29.94  & 33.69  & 0.83  & 0.83  \\
SkT9 & 0.160  & -15.86  & 234.22  & 29.73  & 33.82  & 0.83  & 0.83  \\
SkT9a & 0.160  & -15.86  & 234.22  & 29.73  & 33.82  & 0.83  & 0.83  \\
SKX & 0.155  & -16.03  & 269.76  & 31.07  & 33.40  & 0.99  & 0.75  \\
SKXm & 0.159  & -16.03  & 238.37  & 31.21  & 32.07  & 0.97  & 0.75  \\
Skxs15 & 0.161  & -15.73  & 200.01  & 31.83  & 34.95  & 0.97  & 0.94  \\
SLy0 & 0.160  & -15.95  & 229.00  & 31.95  & 47.10  & 0.70  & 0.80  \\
SLy1 & 0.160  & -15.96  & 229.10  & 31.95  & 47.06  & 0.70  & 0.80  \\
SLy2 & 0.161  & -15.96  & 230.86  & 32.04  & 47.49  & 0.70  & 0.80  \\
Sly230b & 0.160  & -15.95  & 230.84  & 32.04  & 45.99  & 0.69  & 0.80  \\
SLy3 & 0.160  & -15.95  & 228.96  & 31.95  & 45.30  & 0.70  & 0.80  \\
SLy4 & 0.160  & -15.95  & 230.84  & 32.04  & 45.96  & 0.69  & 0.80  \\
SLy5 & 0.161  & -15.96  & 230.77  & 32.05  & 48.18  & 0.70  & 0.80  \\
SLy6 & 0.159  & -15.90  & 229.91  & 31.95  & 47.45  & 0.69  & 0.80  \\
SLy7 & 0.158  & -15.88  & 228.98  & 31.95  & 46.93  & 0.69  & 0.80  \\
SLy8 & 0.160  & -15.95  & 229.18  & 31.96  & 47.16  & 0.70  & 0.80  \\
SLy9 & 0.151  & -15.77  & 229.41  & 31.95  & 54.82  & 0.67  & 0.80  \\
SLy10 & 0.156  & -15.88  & 230.56  & 32.01  & 38.72  & 0.68  & 0.80  \\
QMC600 & 0.174  & -16.40  & 221.21  & 34.65  & 46.81  & 0.81  & 0.61  \\
QMC650 & 0.172  & -16.21  & 221.48  & 33.88  & 53.38  & 0.78  & 0.63  \\
QMC700 & 0.171  & -16.11  & 223.89  & 33.69  & 59.49  & 0.76  & 0.64  \\
QMC750 & 0.171  & -16.21  & 225.98  & 33.96  & 65.10  & 0.74  & 0.65  \\
SV-bas & 0.160  & -15.88  & 234.23  & 30.03  & 32.33  & 0.90  & 0.71  \\
SV-K218 & 0.161  & -15.88  & 217.32  & 29.97  & 34.78  & 0.90  & 0.72  \\
SV-K226 & 0.160  & -15.88  & 224.80  & 29.97  & 34.27  & 0.90  & 0.72  \\
SV-K241 & 0.159  & -15.89  & 241.55  & 30.02  & 30.94  & 0.90  & 0.71  \\
SV-kap20 & 0.160  & -15.88  & 234.08  & 30.03  & 35.52  & 0.90  & 0.83  \\
SV-mas07 & 0.160  & -15.87  & 233.76  & 30.01  & 52.18  & 0.70  & 0.71  \\
SV-mas08 & 0.160  & -15.88  & 233.64  & 30.02  & 40.17  & 0.80  & 0.71  \\
SV-min & 0.161  & -15.89  & 221.55  & 30.65  & 44.85  & 0.95  & 0.93  \\
SV-sym32 & 0.159  & -15.92  & 232.74  & 31.95  & 57.11  & 0.90  & 0.72  \\
SV-sym34 & 0.159  & -15.94  & 233.50  & 33.96  & 80.92  & 0.90  & 0.72  \\
SV-tls & 0.160  & -15.87  & 234.32  & 30.04  & 33.16  & 0.90  & 0.71  \\
T11 & 0.161  & -15.99  & 229.46  & 31.97  & 49.45  & 0.70  & 0.80  \\
T12 & 0.161  & -15.98  & 229.73  & 31.98  & 49.37  & 0.70  & 0.80  \\
T13 & 0.161  & -15.98  & 229.83  & 31.99  & 49.53  & 0.70  & 0.80  \\
T14 & 0.161  & -15.97  & 229.79  & 31.98  & 49.47  & 0.70  & 0.80  \\
T15 & 0.161  & -15.98  & 229.48  & 31.97  & 49.63  & 0.70  & 0.80  \\
T16 & 0.161  & -15.99  & 229.71  & 31.98  & 49.44  & 0.70  & 0.80  \\
T21 & 0.161  & -16.00  & 228.97  & 31.94  & 49.74  & 0.70  & 0.80  \\
T22 & 0.161  & -16.00  & 229.18  & 31.95  & 49.54  & 0.70  & 0.80  \\
T23 & 0.161  & -15.99  & 229.35  & 31.96  & 49.57  & 0.70  & 0.80  \\
T24 & 0.161  & -15.99  & 229.52  & 31.97  & 49.84  & 0.70  & 0.80  \\
T25 & 0.161  & -15.97  & 230.24  & 32.01  & 49.14  & 0.70  & 0.80  \\
T26 & 0.161  & -15.95  & 230.33  & 32.01  & 48.77  & 0.70  & 0.80  \\
T31 & 0.161  & -16.00  & 229.32  & 31.96  & 49.73  & 0.70  & 0.80  \\
T32 & 0.161  & -16.00  & 229.06  & 31.95  & 50.25  & 0.70  & 0.80  \\
T33 & 0.161  & -16.00  & 229.47  & 31.97  & 49.64  & 0.70  & 0.80  \\
T34 & 0.161  & -16.00  & 229.05  & 31.95  & 50.06  & 0.70  & 0.80  \\
T35 & 0.161  & -15.98  & 230.12  & 32.00  & 49.60  & 0.70  & 0.80  \\
T36 & 0.161  & -15.97  & 229.66  & 31.98  & 49.05  & 0.70  & 0.80  \\
T41 & 0.162  & -16.04  & 230.24  & 32.01  & 50.62  & 0.71  & 0.80  \\
T42 & 0.162  & -16.03  & 230.55  & 32.02  & 50.74  & 0.70  & 0.80  \\
T43 & 0.162  & -16.02  & 230.88  & 32.04  & 50.62  & 0.70  & 0.80  \\
T44 & 0.161  & -16.00  & 229.47  & 31.97  & 50.04  & 0.70  & 0.80  \\
T45 & 0.161  & -16.00  & 229.14  & 31.95  & 49.63  & 0.70  & 0.80  \\
T46 & 0.161  & -15.98  & 230.46  & 32.02  & 49.96  & 0.70  & 0.80  \\
T51 & 0.162  & -16.03  & 230.73  & 32.03  & 50.73  & 0.70  & 0.80  \\
T52 & 0.161  & -16.03  & 228.94  & 31.94  & 50.64  & 0.70  & 0.80  \\
T53 & 0.161  & -16.00  & 229.40  & 31.97  & 50.01  & 0.70  & 0.80  \\
T54 & 0.161  & -16.01  & 229.26  & 31.96  & 50.25  & 0.70  & 0.80  \\
T55 & 0.161  & -16.01  & 228.95  & 31.94  & 50.20  & 0.70  & 0.80  \\
T56 & 0.161  & -15.99  & 229.87  & 31.99  & 50.13  & 0.70  & 0.80  \\
T61 & 0.162  & -16.05  & 230.27  & 32.01  & 50.81  & 0.71  & 0.80  \\
T62 & 0.162  & -16.05  & 230.17  & 32.00  & 50.34  & 0.71  & 0.80  \\
T63 & 0.162  & -16.04  & 230.34  & 32.01  & 51.09  & 0.70  & 0.80  \\
T64 & 0.162  & -16.01  & 231.00  & 32.04  & 50.54  & 0.70  & 0.80  \\
T65 & 0.162  & -16.02  & 230.73  & 32.03  & 50.54  & 0.70  & 0.80  \\
T66 & 0.161  & -16.00  & 229.28  & 31.96  & 50.28  & 0.70  & 0.80  \\
\end{supertabular}
\end{center}

\subsection{The symmetry potential}
%
%
%

The symmetry potential $V_\text{sym}$ is also named the Lane potential, which equals the difference between the neutron potential and the proton potential,
\begin{eqnarray} \label{vlane}
\begin{aligned}
    V_\text{Lane} (\rho,p) &= \frac{ V_\text{n} - V_\text{p} }{ 2 \delta } \\
    &= 2 A_\text{sym} \frac{\rho}{ 
        \rho_0 } + 2 B_\text{sym} (\frac{\rho}{\rho_0})^\eta \\
    &\quad + \hbar^2 D_0(\frac{3\pi^2}{2}\rho)^{2/3}\rho   + 2 D_0 m \rho E_k\\
    &= V_{sym}^{loc}+2 D_0 m \rho E_k.
\end{aligned}
\end{eqnarray}
Here, $V_{sym}^{loc}=2 A_\text{sym} \frac{\rho}{\rho_0 } + 2 B_\text{sym} (\frac{\rho}{\rho_0})^\eta + \hbar^2 D_0(\frac{3\pi^2}{2}\rho)^{2/3}\rho $.

To quantitatively understand the momentum dependence and density dependence of $V_\text{Lane}$ on the HICs observables, we investigate the $V_\text{Lane}(\rho,p)$ for two typical Skyrme interaction parameter sets.
They are SkM* and SLy4. The reasons for choosing these two Skyrme interaction parameter sets are as follows. First, the incompressibility ($K_0$), the symmetry energy coefficient ($S_0$) and the isoscalar effective mass ($m_s^*$), should be in a reasonable and commonly accepted range, i.e., $K_0=230\pm 20$ MeV, $S_0=32\pm 2$ MeV, and $m^*_s/m=0.7\pm 0.1$. Second, the parameter sets have different signs of the effective mass splitting, $\delta m^*_{np}=(m_n^*-m_p^*)/m>0$ or $<0$. 
The set of SLy4 \cite{chabnate} has $\delta m^*_{np}<0$ (or $f_I>0$) in neutron rich matter, and the slope of symmetry energy $L$ is 46 MeV. The set SkM* has $\delta m^*_{np}>0$ (or $f_I<0$) and $ L = 46 $ MeV. For convenience, the values of the nuclear matter parameters of SkM*, SLy4 are listed in Table~\ref{tbl402}.
%
%
%
\begin{table}[htbp]
\centering
\caption{Nuclear matter parameters of SLy4 and SkM*. The parameters $E_0$, $K_0$, $S_0$, $L$ are in MeV, and $\rho_0$ is fm$^{-3}$.}
\label{tbl402}
\begin{tabular}{cccccccc} \\ \hline
Para. & $ \rho_0 $ & $ E_0 $ & $ K_0 $ & $ S_0 $ & $ L  $ & $ m_s^*/m $ & $ m_v^*/m $ \\ \hline
SLy4~ & ~0.160~ & ~-15.97~ & ~230~ & ~32~ & ~46~ & ~0.69~ & ~0.80 \\
SkM*~ & ~0.160~ & ~-15.77~ & ~217~ & ~30~ & ~46~ & ~0.79~ & ~0.65 \\ \hline
\end{tabular}%

\end{table}
\begin{figure}[!ht]
    \centering
    \includegraphics[width=\linewidth]{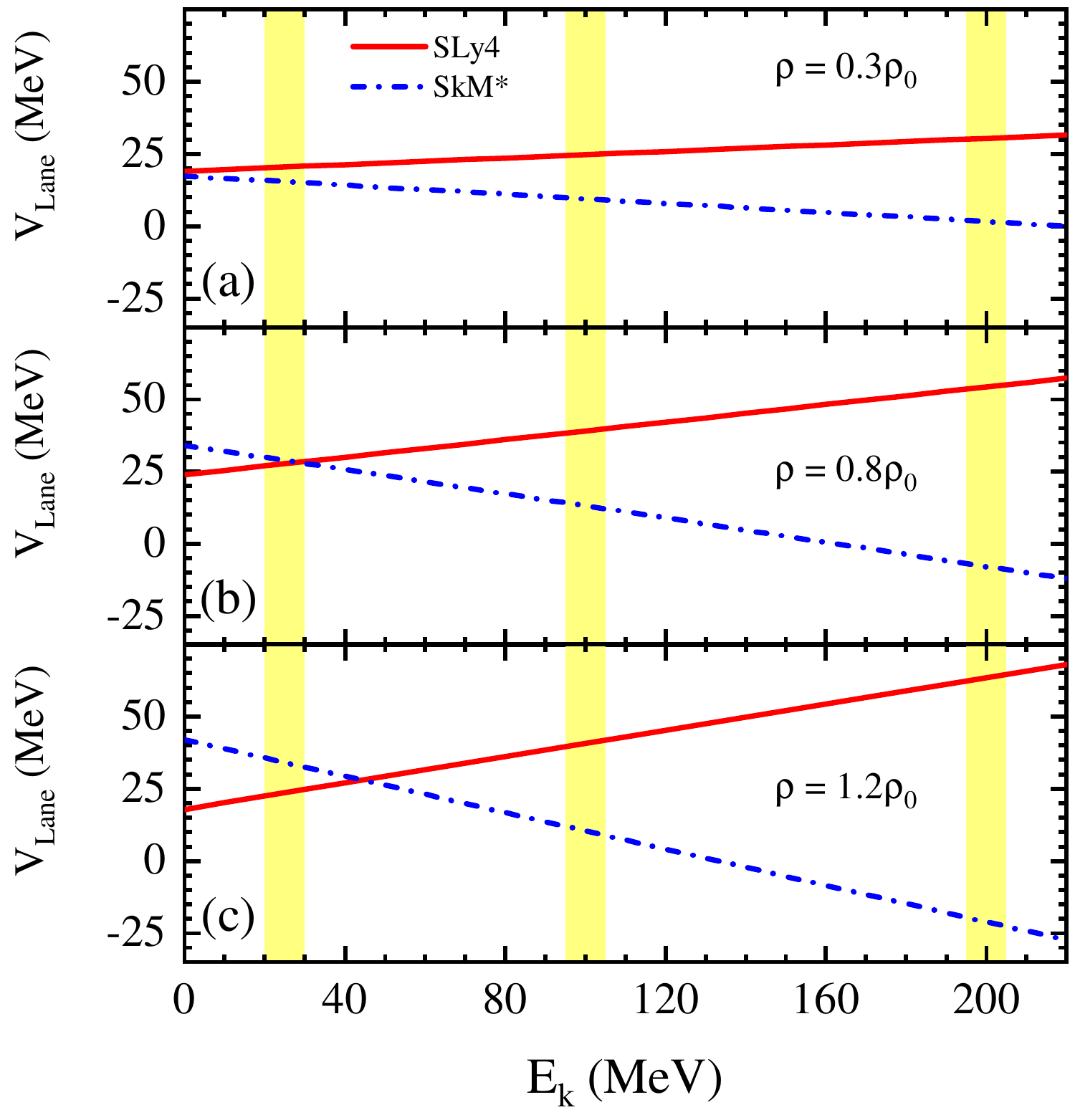}
    \caption{The Lane potential $V_\text{Lane}$ as functions of kinetic energy $ E_k $ at density $ \rho = 0.3 \rho_0,~0.8 \rho_0 \text{ and } 1.2 \rho_0 $.}
    \label{fig2-usym}
\end{figure}

In Figure~\ref{fig2-usym}, we present the $V_\text{Lane}$ as functions of kinetic energy for cold nuclear matter with isospin asymmetry $\delta=0.2$. The top panel is the results at $\rho=0.3\rho_0$, the middle panel is for 0.8$\rho_0$, and the bottom panel is for 1.2$\rho_0$. The $V_\text{Lane}$ increases (decreases) with the kinetic energy increasing for $\delta m^*_{np}<0$ ($\delta m^*_{np}>0$). They will influence the neutron to proton yield ratios Y(n)/Y(p) as a function of kinetic energy in HICs according to the following relationship, 
\begin{equation} \label{ynyp-vlane}
\begin{aligned}
    \frac{\text{Y(n)}}{\text{Y(p)}} &\propto \exp \left( \frac{ \mu_n - \mu_p}{ T }\right) \\
   &= \exp \left[ \frac{ 2\left( V_{sym}^{loc} +2D_0m\rho E_k \right) \delta }{ T } 
        \right].
\end{aligned}
\end{equation}
The above relation can be obtained by the statistical and dynamical model \cite{Tsang2001prl, Tsang2001prc, Ono2003, Das1981, George1987, Botvina2002}. $T$ is the temperature of emitting source, $\mu_n$ and $\mu_p$ are the chemical potentials of neutrons and protons, respectively. 
Thus, one can expect that the larger the Lane potential, the larger the neutron to proton yield ratios. Similar effects on the triton to $^3$He are also expected \cite{chajecki1402.5216}, i.e.,
\begin{equation} \label{ytyhe3-vlane}
\begin{aligned}
    \frac{\text{Y(t)}}{\text{Y($^3$He)}} &\propto \exp \left( \frac{ \mu_\text{t} - \mu_{ ^3\text{He} } }{ T }\right)\approx\exp  \left( \frac{ \mu_n - \mu_{p}}{ T }\right) \\
   &= \exp \left[ \frac{ 2 \left( V_\text{sym}^\text{loc}+2D_0m\rho E_k \right) \delta }{ T } 
        \right].
\end{aligned}
\end{equation} 
In addition, one can expect that the slopes of Y(n)/Y(p) ratios with respect to the $E_k$ will be different with effective mass splitting according to Eq. \eqref{ynyp-vlane}, and the similar behaviour is also expected for Y(t)/Y($^3$He).

\subsection{Y(n)/Y(p) and Y(t)/Y(\texorpdfstring{$^3$}{3}He)}

To see the effects of the effective mass splitting on the HICs observables, such as Y(n)/Y(p) and Y(t)/Y($^3$He), we perform the simulation of $^{86} $Kr+$ ^{208} $Pb at the beam energy from $ E_\text{beam} = 25 \text{A MeV} $ to 200A MeV. In the calculations, the impact parameter $ b = 1 $ fm, the number of events is 100, 000. The dynamical evolution time is up to 400 fm/$c$.
\begin{figure}[!b]
    \centering
    \includegraphics[width=\linewidth]{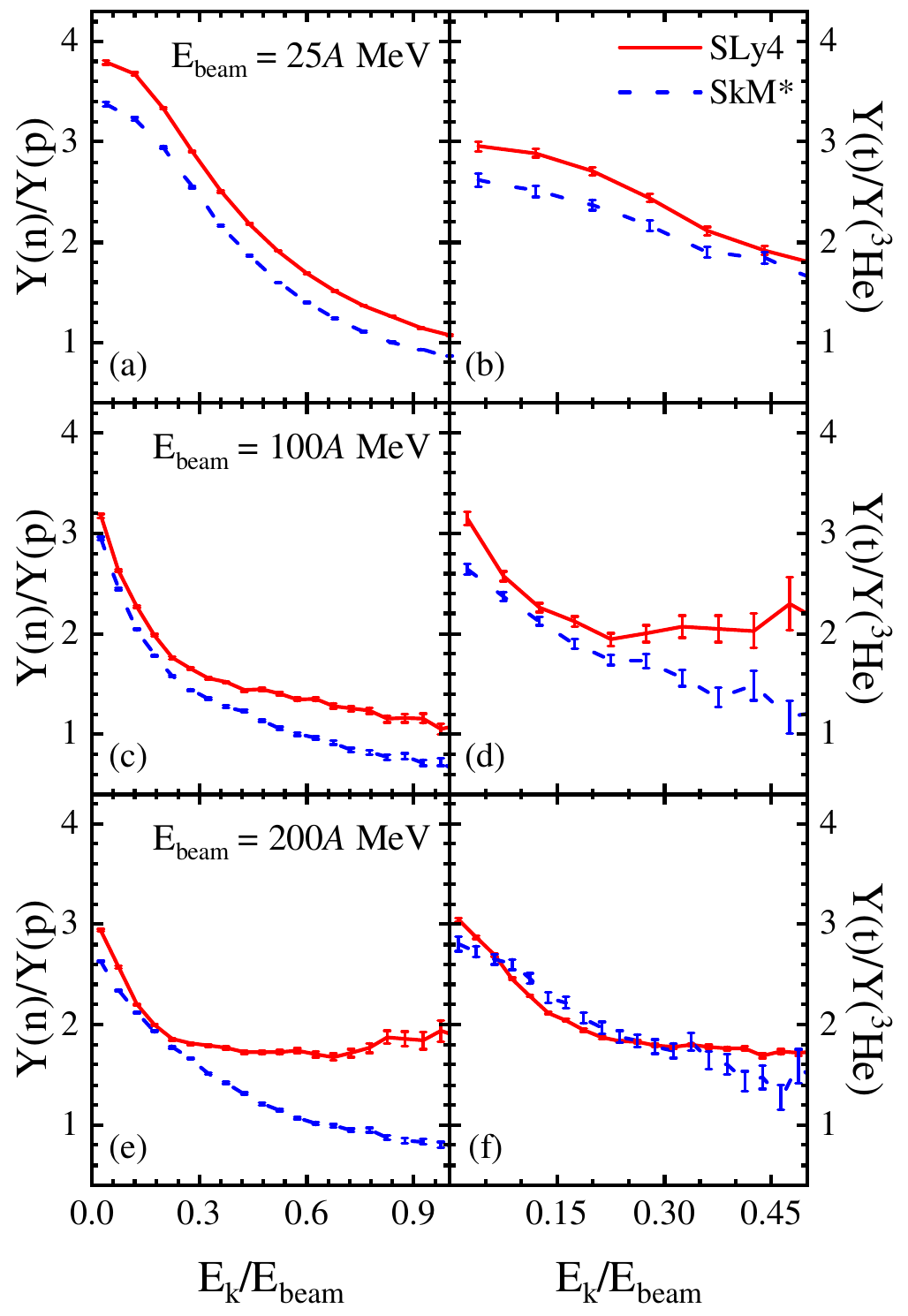}
    \caption{The yield ratio of Y(n)/Y(p) and Y(t)/Y($^3$He) as a function of the normalized nucleon center of mass energy $E_k/E_\text{beam}$ at the beam energy $ E_\text{beam} = 25, 100 \text{ and } 200 \text{A MeV} $.}
    \label{fig3v9}
\end{figure}

The left panels of Figure~\ref{fig3v9} show the Y(n)/Y(p) ratios as functions of the normalized nucleon center of mass energy, i.e., $E_k/E_\text{beam}$. The errors of Y(n)/Y(p) are statistical uncertainties, which are obtained by the error propagation formula from the errors of Y(n) and Y(p). By using $E_k/E_\text{beam}$, the shape of Y(n)/Y(p) as functions of kinetic energy can be compared and understood on a similar scale for different beam energies. The red lines correspond to the results obtained with SLy4 ($m_n^*<m_p^*$) and the blue lines are for SkM* ($m_n^*>m_p^*$). Our calculations show that both the Y(n)/Y(p) ratios obtained with SLy4 and SkM* decrease with the nucleon kinetic energy increasing due to the Coulomb effects. Furthermore, the Y(n)/Y(p) ratios obtained with SLy4 ( $m_n^*<m_p^*$) are larger than that obtained with SkM* ($m_n^*>m_p^*$). At the beam energy of 200A MeV, a flatter Y(n)/Y(p) dependence on the nucleon kinetic energy can be observed for SLy4. This is because SLy4 has stronger Lane potentials at high kinetic energy and enhance neutron emissions at high nucleon energies. 

In more detail, the difference of Y(n)/Y(p) between the SLy4 ($m_n^*<m_p^*$) and SkM*($m_n^*>m_p^*$) keeps a constant value with nucleons kinetic energy at 25A MeV and increases with nucleons kinetic energy at the beam energy greater than 100A MeV. It can be understood from the symmetry potential as shown in Fig.~\ref{fig2-usym}. At 25A MeV, the system is less compressed and less excited than that at 100 or 200A MeV, and most of the emitted nucleons are from the low density region. The corresponding symmetry potentials for SLy4 and SkM* varies weakly as a function of kinetic energy (see Fig. \ref{fig2-usym}(a)). Thus, one can expect that the difference of Y(n)/Y(p) between the SLy4 ($m_n^*<m_p^*$) and SkM* ($m_n^*>m_p^*$) is small and changes weakly as the kinetic energy increases. At a beam energy greater than 100A MeV, the system can be compressed to higher densities, where the magnitude of the splitting increases with the kinetic energy as shown in Fig.~\ref{fig2-usym} (b) and (c). 

The right panels of Figure~\ref{fig3v9} show the Y(t)/Y($^3$He) ratios as functions of the normalized nucleon center of mass energy, i.e., $E_k/E_{b}$. Similar to Y(n)/Y(p), the Y(t)/Y($^3$He) ratios also show a sensitivity to the effective mass splitting. The reason can also be understood from Eq.(\ref{ytyhe3-vlane}). At the beam energy of 200A MeV, the sensitivity of the Y(t)/Y($^3$He) ratios to kinetic energy becomes weak which may be due to the cluster effects and stronger non-equilibrium effects than that at lower beam energies.

Furthermore, Fig.~\ref{fig3v9} also shows that the Y(n)/Y(p) ratio has an exponential decreasing 
approximately with respect to $E_k/E_\text{beam}$ in the range of $ 0.3 \leqslant E_k/E_\text{beam} \leqslant 1.0 $. For t/$ ^3 $He ratios, a similar behaviour can be found in $ 0.2 \leqslant E_k/E_\text{beam} \leqslant 0.5 $. The reason of $E_k/E_\text{beam}\leqslant 0.5$ for Y(t)/Y($^3$He) ratios is the kinetic energy per nucleon for the emitted tritons or $^3$He is about one half of the beam energy. According to Eq.(\ref{ynyp-vlane}) and Eq.(\ref{ytyhe3-vlane}), the character of the exponential decreasing behaviour reflects the emitted nucleons are in equilibrium in momentum space and can be described by the slopes of $\ln $[Y(n)/Y(p)] or $\ln $[Y(t)/Y($^3$He)]. The slopes of $\ln $[Y(n)/Y(p)] or $\ln $[Y(t)/Y($^3$He)] read, 
\begin{equation} \label{230322224103}
\begin{aligned}
    S_\text{n/p} &= \frac{\partial \ln[\text{Y(n)/Y(p)} ]}{\partial E_k}=4D_0 m\delta \rho/T, \\
    S_{\text{t/}^3\text{He}} &= \frac{\partial \ln[\text{Y(t)/Y(} ^3 \text{He)} ]}{\partial E_k}=4D_0 m\delta\rho/T.
\end{aligned}
\end{equation}
They are directly related to the $D_0$.

In the following analysis, we do linear fit of the $\ln $[Y(n)/Y(p)] and $\ln $[Y(t)/Y($^3$He)] as
\begin{equation} \label{lnnp-slb}
    \ln \left[\frac{ \text{Y(n)} }{ \text{Y(p)} } \right] = S_\text{n/p} \frac{E_k}{E_\text{beam}} + b_0^\text{n/p}
\end{equation}
in the range of $0.3 \leqslant E_k/E_\text{beam} \leqslant 1.0$ and
\begin{equation} \label{lnthe3-slb}
    \ln \left[\frac{ \text{Y(t)} }{ \text{Y($^3$He)} } \right] = S_\text{t/3He} \frac{E_k}{E_\text{beam}} + b_0^\text{t/$^3$He}
\end{equation}
in the range of $0.2 \leqslant E_k/E_\text{beam}\leqslant 0.5$, respectively, to get the slopes of $ S_\text{n/p} $ ($ S_{ \text{t/} ^3 \text{He}} $), and intercepts of $ b_0^{n/p} $ ($ b^{t/^3\text{He}}_0 $). To describe the goodness of this linear fit of the ln(Y(n)/Y(p)) and ln(Y(t)/Y($^3$He)), we present the coefficients of determination $R^2$~\cite{Alessandro2008} in Table \ref{tbl03}. 
\begin{table}[!h]
    \centering
    \caption{The coefficients of determination $R^2$ for the linear fit of the ln(Y(n)/Y(p)) and ln(Y(t)/Y($^3$He)).}
    \label{tbl03}
    \begin{tabular}{ccccc} \hline
        \multirow{2}[4]{*}{$R^2$} & \multicolumn{2}{c}{ln(Y(n)/Y(p))} & \multicolumn{2}{c} 
            {ln(Y(t)/Y($^3$He))} \bigstrut\\
        \cline{2-5}  & SLy4 & SkM* & SLy4 & SkM* \bigstrut\\ \hline
        25A MeV & 0.98767 & 0.98431 & 0.99073 & 0.97503 \bigstrut[t]\\
        100A MeV & 0.97342 & 0.98295 & 0.50164 & 0.84234 \\
        200A MeV \qquad & 0.36945 \qquad & 0.96828 \qquad & 0.79116 \qquad & 0.66351 
            \bigstrut[b]\\ \hline
        \end{tabular}%
\end{table}
\begin{figure}[!ht]
    \centering
    \includegraphics[width=\linewidth]{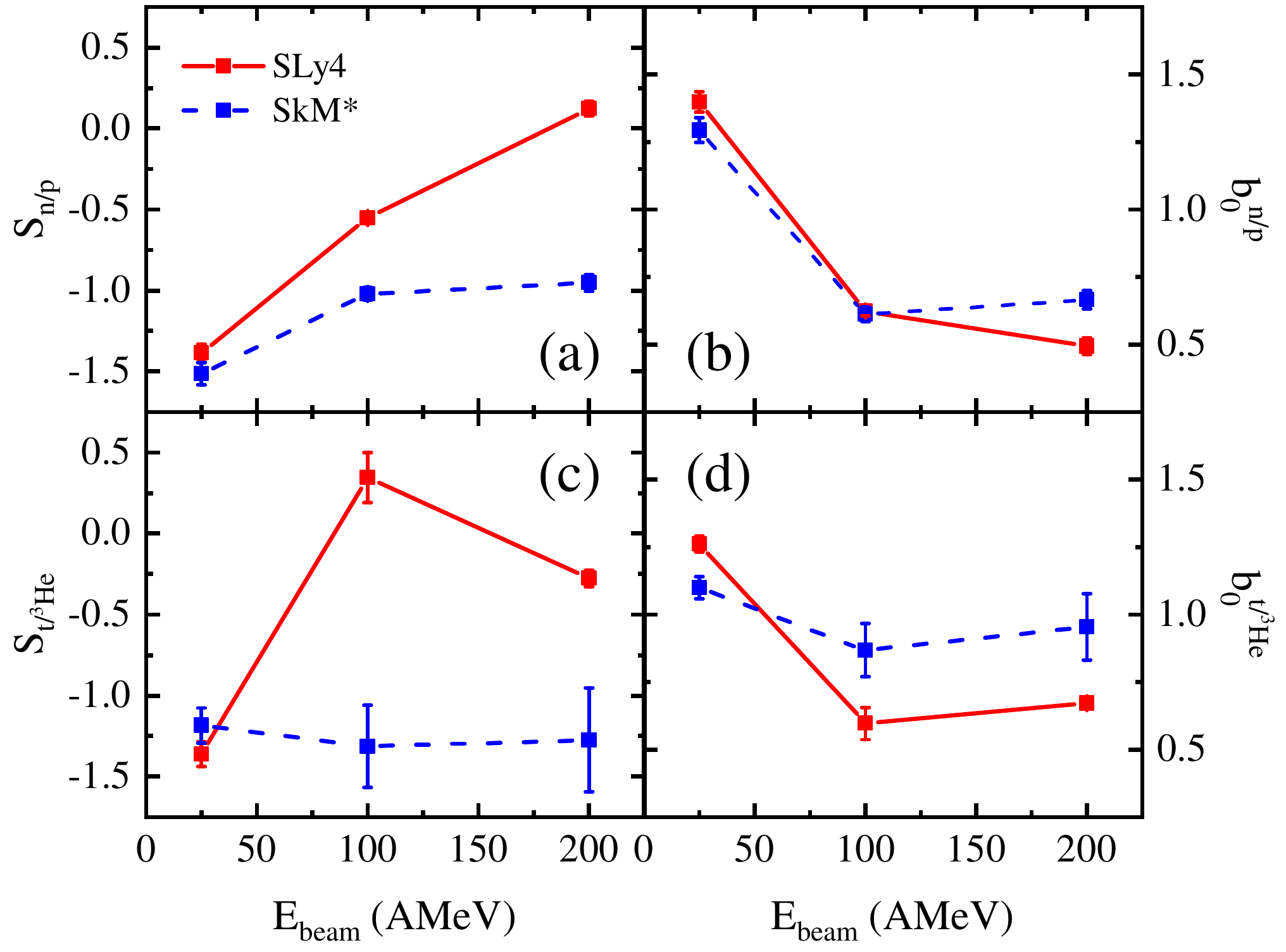}
    \caption{The $ S_X $ and $ b_0^{X} $ as functions of $E_\text{beam}$ for SLy4 and SkM*. Upper panels are $X$= n/p and bottom panels are $X$=t/$^3$He.}
    \label{fig4v10}
\end{figure}
%

%
%

Fig.~\ref{fig4v10} presents $ S_\text{n/p} $ ($ S_{ \text{t/} ^3 \text{He}} $) and $ b_0^\text{n/p} $ ($ b_0^{t/^3\text{He}} $) as functions of beam energy for seeking the optimal energy. Panel (a) is for $ S_\text{n/p} $, and panel (c) is for $ S_{ \text{t/} ^3 \text{He}} $. Our calculations show that the values of $ S_\text{n/p} $ ($ S_{ \text{t/} ^3 \text{He}} $) obtained with SLy4 are higher than that with SkM* except for the beam energy at 25A MeV. In more detail, the impacts of effective mass splitting on $ S_\text{n/p} $ become obvious at the beam energy 200A MeV. For $ S_{ \text{t/} ^3 \text{He}} $, the impacts of the effective mass splitting reach the largest at the beam energy around 100A MeV under the influence of the cluster formation mechanism. For the $ b_0^\text{n/p} $, the calculations show that it weakly depends on the effective mass splitting, except for the value of $ b_0^\text{n/p} $ at $E_{b}=200$A MeV. At this beam energy, the value of $ b_0^\text{n/p} $ obtained with SkM* is larger than that obtained with SLy4. For $ b_0^{t/^3\text{He}} $, the value obtained with SkM* is less than that with SLy4 at the beam energy 25A MeV. At the beam energy larger than 100A MeV, the values of $b_0^{t/^3\text{He}}$ obtained with SkM* are greater than that with SLy4. 


%
%

\section{Summary and outlook} \label{summar}
In summary, we have compiled the 119 Skyrme interaction sets and the corresponding nuclear matter parameters for understanding the correlations among the different nuclear matter parameters. By analyzing the linear correlation coefficient, the strength of the correlation between different nuclear matter parameters is quantitatively obtained. Further, the correlation among different nuclear parameters means the tight constraints of the symmetry energy not only need to know the values of the symmetry energy coefficient $S_0$ and the slope of the symmetry energy $L$, but also need to know the isoscalar effective mass $m_s^*$ and the isovector effective mass $m_v^*$ or the effective mass splitting given the $K_0$ and $E_0$ are well constrained.

To understand the impacts of the effective mass splitting on HICs observables, we simulated the $^{86}$Kr+$^{208}$Pb at the beam energy ranging from 25 MeV to 200 MeV per nucleon. Two observables are analyzed, one is the emitted neutron to proton ratios, and another is the triton to Helium-3 ratios. Our results show that the energy spectra of Y(n)/Y(p) and Y(t)/Y(3He) can be used to distinguish the effective mass splitting, which is consistent with previous studies in Ref. \cite{Zhang2014,Rizzo2004}. Further, we construct the characteristic variables, which are the slope and intercept of the $\ln $[Y(n)/Y(p)] and $\ln $[Y(t)/Y($^3$He)] and can be directly related to the effective mass splitting. 
The largest effects can be found at 200A MeV for (Y(n)/Y(p)), while the largest effects can be found at 100A MeV for (Y(t)/Y($^3$He)). This difference may come from the cluster formation mechanism.

\appendix
\section{Single particle potential}
\label{Vq}

For the Skyrme interaction, the single particle potential in uniform nuclear matter can be written as the summation of the local and nonlocal (momentum dependent) parts as follows,
\begin{equation}
V_q=V_q^\text{loc}+V_{q}^\text{md}.
\end{equation}

Based on the definition of the single particle potential, $V_q$ should be obtained from the derivatives of the net energy $E$ of a system with respect to the number of particles. For the local part, $V_q^\text{loc}$ reads,
\begin{equation} \label{vloc}
\begin{aligned}
    V^\text{loc}_q (\rho, \delta) &= \frac{ \partial u_\text{loc }( \rho, \delta) }{ 
        \partial \rho_q } \\
    &= \alpha \frac{\rho }{ \rho_0 } + \beta  \frac{ \rho^\eta }{ \rho_0^\eta } 
        + (\eta - 1) B_\text{sym} \frac{\rho^\eta}{\rho_0^\eta}\delta^2\\
    & \quad \pm 2 \left(A_\text{sym} \frac{\rho}{\rho_0}+ 
        B_\text{sym} \left( \frac{\rho}{\rho_0} \right) ^{\eta} \right) \delta.
\end{aligned}
\end{equation}
The sign `$+$' is for neutrons, and `$-$' for protons. The nonlocal part of the single particle potential depends not only on the position but also on the momentum, it can be obtained by taking the functional derivative of the energy density with respect to the distribution function $f(r,p)$ , we 
%
\begin{eqnarray} \label{vnonlocal}
\begin{aligned}
    V^\text{md}_q(\rho,\delta,p) &= \frac{\delta u_\text{md}}{\delta 
        f_q} \\
    &= 2(C_0\rho+D_0\rho_q)p^2+ 2\hbar^2C_0\tau+2\hbar^2D_0\tau_q\\
\end{aligned}
\end{eqnarray}
Here, $\tau$ is the kinetic energy density and is the summation of kinetic energy densities of neutrons and protons, i.e., $\tau=\tau_n+\tau_p$. $\tau_q=\frac{3}{5}k_{q,F}^2\rho_q$, with $k_{q,F}=(3\pi^2\rho_q)^{1/3}$.

\bibliography{ref}

\end{document}